\documentclass[journal = jacsat]{achemso}

\usepackage{amsmath,amssymb}
\usepackage{color}

\usepackage{graphicx}

\title{Limiting performance of graphene bilayer sub-terahertz detectors at large induced band gap}

\author{Elena I. Titova}
\affiliation{Center for Photonics and 2D Materials, Moscow Institute of Physics and Technology, Dolgoprudny, 141700, Russian Federation}
\alsoaffiliation{Programmable Functional Materials Lab,
Center for Neurophysics and Neuromorphic Technologies, Moscow, 121205, Russia}
\email{titova.elenet@gmail.com}

\author{Mikhail A. Kashchenko}
\affiliation{Center for Photonics and 2D Materials, Moscow Institute of Physics and Technology, Dolgoprudny, 141700, Russian Federation}
\alsoaffiliation{Programmable Functional Materials Lab,
Center for Neurophysics and Neuromorphic Technologies, Moscow, 121205, Russia}

\author{Andrey V. Miakonkikh}
\affiliation{Valiev Institute of Physics and Technology RAS, 117218, Moscow, Nakhimovsky av. 36/1}

\author{Alexander D. Morozov}
\affiliation{Programmable Functional Materials Lab,
Center for Neurophysics and Neuromorphic Technologies, Moscow, 121205, Russia}

\author{Ivan K. Domaratskiy}
\affiliation{Center for Photonics and 2D Materials, Moscow Institute of Physics and Technology, Dolgoprudny, 141700, Russian Federation}

\author{Sergey S. Zhukov}
\affiliation{Center for Photonics and 2D Materials, Moscow Institute of Physics and Technology, Dolgoprudny, 141700, Russian Federation}

\author{Vladimir V. Rumyantsev}
\affiliation{Institute for Physics of Microstructures, Russian Academy of Sciences, 603950 Nizhny Novgorod, Russia}

\author{Sergey V. Morozov}
\affiliation{Institute for Physics of Microstructures, Russian Academy of Sciences, 603950 Nizhny Novgorod, Russia}

\author{Kostya S. Novoselov}
\affiliation{Institute for Functional Intelligent
Materials, National University of Singapore, Singapore, 117575, Singapore}

\author{Denis A. Bandurin}
\affiliation{Department of Materials Science and Engineering, National University of Singapore, Singapore, Singapore}

\author{Dmitry A. Svintsov}
\affiliation{Center for Photonics and 2D Materials, Moscow Institute of Physics and Technology, Dolgoprudny, 141700, Russian Federation}
\email{svintcov.da@mipt.ru}

\date{November 2024}

\begin{document}

\maketitle
\date{\today}

\begin{abstract}
Electrically induced $p-n$ junctions in graphene bilayer (GBL) have shown superior performance for detection of sub-THz radiation at cryogenic temperatures, especially upon electrical induction of the band gap $E_g$. Still, the upper limits of responsivity and noise equivalent power (NEP) at very large $E_g$ remained unknown. Here, we study the cryogenic performance of GBL detectors at $f=0.13$ THz by inducing gaps up to $E_g \approx 90$ meV, a value close to the limits observed in recent transport experiments. High value of the gap is achieved by using high-$\kappa$ bottom hafnium dioxide gate dielectric. The voltage responsivity, current responsivity and NEP optimized with respect to doping do not demonstrate saturation with gap induction up to its maximum values. The NEP demonstrates an order-of-magnitude drop from $\sim450$ fW/Hz$^{1/2}$ in the gapless state to $\sim30$ fW/Hz$^{1/2}$ at the largest gap. At largest induced band gaps, plasmonic oscillations of responsivity become visible and important for  optimization of sub-THz response.
\end{abstract}

A remarkable property of two-dimensional materials lies in the possibility of all-electrical induction of $p-n$ junctions with the aid of gating~\cite{Ryzhii_PNJ}. Such structures acquire the possibility to rectify (detect) electromagnetic radiation in the terahertz and infrared ranges~\cite{Castilla2019,Castilla_plasmonic_antenna,MATBG_split_gate}. Importantly, such method of junction creation does not introduce dopants and preserves high mobility and fast response~\cite{Castilla_plasmonic_antenna}. 

Electromagnetic optimization of 2d $p-n$ junctions in terms of matching with terahertz antennas has led to great results in terms of high responsivity and low noise equivalent power~\cite{Lisauskas_antenna,Drew_antenna,Koppens_Wireless}. Complementary to the electromagnetic matching, boosting the intrinsic rectifying capability of the 2d $p-n$ junctions is an important way to detector enhancement~\cite{Gayduchenko_TFET}. The rectifying properties of $p-n$ junctions are traced down to the presence of energy barriers for the charge carriers. It implies that junctions with gapless materials are less efficient for rectification, as compared to the gapped ones. This statement was experimentally proved recently~\cite{Titova_BLG} using bilayer graphene (BLG), a unique material with electrically-tunable band gap $E_g$. It was shown that  cryogenic sub-THz responsivity of $p-n$ junctions in BLG displays a $3$-fold growth with changing the gap from zero to $\sim 20$ meV at $T\approx 20$ K. Further gap enhancement was obstacled by the non-optimal capacitance of the gates. This left the question about the limiting performance of BLG sub-THz rectifiers with large induced gaps open.

The present paper aims to study the maximum responsivity and minimum noise equivalent power (NEP) of graphene bilayer sub-THz detectors at maximum achievable band gaps in this material. Theoretically, the gap value is limited by $\gamma_1 \approx 0.37$ eV, the interlayer hopping integral. Most electrical studies of high-quality encapsulated structures report the upper limit of $E_g$ slightly exceeding $0.1$ eV, as inferred from measurements of temperature-activated resistivity~\cite{Icking_spectroscopy,Koppens_VHE_BLG,Icking_steep}. We achieve similar values of transport gap and proceed to study the terahertz rectification in such structures.

To increase the efficiency of gap induction in BLG, we have replaced the conventionally used boron nitride back gate dielectric with high-$\kappa$ hafnium dioxide grown by atomic layer deposition. The thickness of HfO$_2$ layer is 27 nm and its dielectric constant $\kappa \sim 10$. The layer was deposited on $p$-doped silicon wafer with $12$ Ohm $\cdot$ cm bulk resistivity. The latter acted as a back gate transparent at sub-THz frequencies. Atop of HfO$_2$, we have transferred an encapsulated stack comprised of bottom hBN (1.6 nm), BLG, and top hBN (30 nm). The edges of BLG, being naturally almost parallel, were not exposed to any etching. This prevented the emergence of edge channels that could limit the insulting resistivity of the device~\cite{Stampfer_QDs,Domaratskiy_NatEdge,Bandurin_EdgeChannels}. Finally, Cr/Au contacts and gates were deposited to define the split-gate configuration suitable for $p-n$ junction induction. We intentionally retained a large gap between metal gates, equal nominally to the half of the channel length, to increase the coupling of radiation to the induced junctions.

\begin{figure}
    \centering
    \includegraphics[width=1\linewidth]{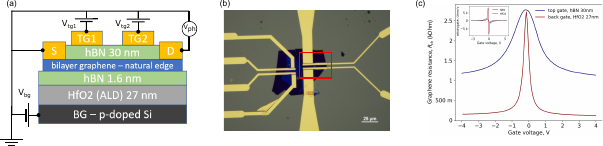}
    \caption{(a) Schematic of the split-gate bilayer graphene terahertz detector. (b) Optical micro-photograph of the device with its contacts. The characteristics of the device on the right were measured. (c) Resistance of BLG channel recorded as a function of back gate voltage (red) and top gate voltage (blue), demonstrating superior resistance tunability through the high-$\kappa$ stack. Insert: $dR_{ch}/dV_{gate}$ as a function of back (red) and top (blue) gate voltages.}
    \label{fig1}
\end{figure}

The final structure and measurement scheme are shown in Fig.~\ref{fig1}, (a) and (b). The device under study  (highlighted with a red rectangle in Fig.~\ref{fig1}, (b)) contained a terahertz bow-tie antenna connected to the channel. To measure the graphene resistance, we applied an alternating current of about $I_{\rm sd} = 25$ nA to the drain contact at a frequency $f_I=83$ Hz, whereas the THz photoresponse emerged and was measured without current bias. The dependence of graphene resistance $R$ on the voltage on the top ($V_{tg}$) and back ($V_{bg}$) gates is shown in Fig.~\ref{fig1} (c). The slope of $R(V_{bg})$-curves more than twice exceeds the slope of $R(V_{tg})$-curves, confirming the large dielectric constant of HfO$_2$.  

\begin{figure}
    \centering
    \includegraphics[width=1\linewidth]{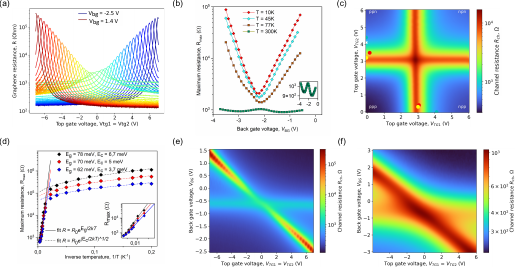}
    \caption{Electrical characterization of graphene bilayer THz detector
    (a) Channel resistance as a function of top gate voltage at fixed back gate voltages. (b) Maximum BLG resistance reached upon top gate scanning (charge neutrality resistance) versus back gate voltage at different temperatures. Insert shows the magnified view of maximum room-temperature resistance. (c) Cryogenic BLG resistance as a function top gate voltages $V_{tg1}$ and $V_{tg2}$ at fixed back gate voltage $V_{bg} = -1.42$  V. (d) Arrhenius plots of the charge neutrality resistance at different back gate voltages (controlling the band gap). The dots indicate experimental data. Solid lines show the fits according to the conventional activation law $R\propto e^{E_g/2kT}$ (high temperatures) and Shklovsky-Efros law $R\propto \exp[(E_c/2kT)^{1/2}]$ (low temperatures) (e,f) Color map of graphene resistance vs back gate voltage $V_{bg}$ and top gate voltage $V_{tg1=tg2}$ recorded for cryogenic (e) and room (f) temperatures.}
    \label{fig2}
\end{figure}

Room-temperature electrical characterization of gate-dependent resistivity $R(V_{tg},V_{bg})$ shows just a moderate effect of gap induction [see Fig.~\ref{fig2}, (b,f)]. It manifests as an increase in resistivity at the charge neutrality point (CNP) with increasing the back gate voltage $V_{bg}$. The effect is largely masked by resistivity reduction in non-gated sections of the channel under the same back gate polarity. Still, the evidence of increased resistivity at room temperature point to the induction of quite large gaps in the channel.

Efficient gating by back gate with HfO$_2$ stack becomes more apparent in cryogenic measurements of $R(V_{tg},V_{bg})$. The maps of $R(V_{tg},V_{bg})$-dependences demonstrate a well-known pattern with CNP resistance at the map diagonal increasing in its top left and bottom right corners, as shown in Fig.~\ref{fig2} (e,f). A resistance maximum at the horizontal line of such a map corresponds to the neutrality condition of an area in the middle of the channel controlled by the back gate. Its influence becomes less pronounced at low temperatures, where the resistance depends on the induced gap $E_g$ exponentially, while the magnitude of $E_g$ in the single-gated region is small. 

Still, the central region should be taken into account when considering $p-n$ junctions in the channel. Fig.~\ref{fig2} (c) shows graphene resistance as a function of two top gates at fixed negative back gate voltage $V_{bg} = -1.42 V$. The resistance increase in the first quadrant corresponds to the opposite doping of the central region in relation to the rest of the channel. In the second and fourth quadrants, $p-n$ and $n-p$ junctions are induced by opposite voltages on the top gates $V_{tg1}$ and $V_{tg2}$, and the second quadrant corresponds to the uniform channel doping.

Variable-temperature measurements of CNP resistance display a more interesting pattern. At 100 K $< T < $ 300 K, the resistance shows a conventional Arrhenius-type behavior well fitted by $R = R_0 \exp\left( E_g / 2kT \right)$. The gaps obtained from such fitting reach $80$ meV. At lower temperatures, the resistance growth with $1/T$ slows down. However, the growth of resistance does not slow down with the back gate, and there are no traces of resistance saturation with increasing $V_{bg}$. A possible explanation of such low-temperature scaling lies in variable-range Shklovsky-Efros hopping between charge puddles formed by defects on HfO$_2$ interface. Indeed, the low-$T$ resistance is well fitted by the theoretically appropriate dependence $R = R_0 \exp\left( [E_c / 2kT]^{p} \right)$ with $p=1/2$~\footnote{Previous studies~\cite{Herrero_Large_D,Zou_fluctuations} used $p=1/3$ which corresponds to the model of variable range hopping in two dimensions. Our $p=1/2$ corresponds to the account of Coulomb charging of the puddles upon hopping transport~\cite{Efros_CoulombGap}. Strictly speaking, our resistance data are best described by $p\approx0.06$.
} and a lower characteristic energy $E_c$. Its value reaches 7 meV at the largest available back gate voltage. It is interesting that the dependence of CNP resistance on the displacement field $D$ is still exponential in the Shklovsky-Efros regime, as displayed in Fig.~\ref{fig2} (d). This falls in qualitative agreement with recent results of Ref.~\cite{Shklovsky_Efros_2022}, where a strong gap dependence of characteristic energy $E_C \propto E_g^{55/9}$ was predicted.

After obtaining strong experimental evidence of large induced band gap, we proceed to the sub-terahertz response measurements of the graphene transistor. We focused sub-THz radiation with $f=0.13$ THz generated by an avalanche diode with nominal power $p_0\approx 10$ mW to the sample using TPX lens. A set of sub-THz attenuators passing down to $k = 3\times 10^{-4}$ fraction of the net power was used to keep the device in the linear regime. The rectified photovoltage $V_{\rm ph}$ was measured between the source and drain using the lock-in technique at the modulation frequency $f_{\rm mod} = 14$ Hz of the incident radiation. The magnitude of responsivity was varied by alternating the split top gate voltages $V_{tg1}$ and $V_{tg2}$, which resulted in the $p-n$ or $n-p$ junction formation along the channel. Special precautions were taken to avoid the photovoltage generation due to unintentional asymmetry of the sub-THz illumination. To this end, we have recorded the maps of $V_{\rm ph}$ upon $x-y$ scanning of the sub-THz source, and selected the position where the signals in $p-n$ and $n-p$ configurations are the opposite. Still, a slightly asymmetric illumination only resulted in a decrease in photoresponse, without any qualitative changes in the results.

%Optical studies report higher gap values, which is explained by a different effect of built-in disorder on optical absorbance and static conduction. Indeed, the transport channels formed by disorder are in parallel with bulk thermally-activated transport, thus effectively shorting the source and drain in dc measurements with high barrier. In optical studies, such states produce a small additive contribution to absorption, resulting in weak tails below the absorption edge. Despite the differences in gap values obtained with these different methods, we're dealing with physically identical samples of bilayer graphene with very large induced interlayer potential.

\begin{figure}
    \centering
    \includegraphics[width=1\linewidth]{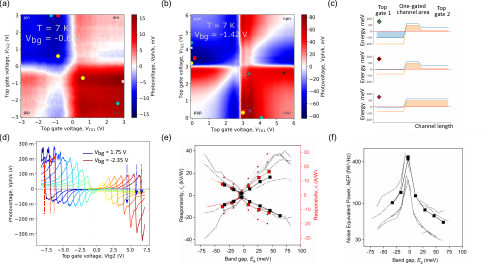}
    \caption {Terahertz photoresponce of BLG detector with induced $p-n$ junctions. (a,b) Color map of BLG sub-THz photovoltage as a function top gate voltages $V_{tg1}$ and $V_{tg2}$ controlling the density at two sides of the channel. $k$ is the attenuation coefficient of the THz power.  Panel (a) corresponds to the nominally gapless state, panel (b) -- to the state with large induced band gap. Red dots on the color maps represent the optima of photovoltage  $V^{\max}_{\rm ph}$, cyan dots -- the optima of photocurrent $I^{\max}_{\rm ph}$, yellow -- the optima of photovoltage per THz power dissipated in the channel $(V_{\rm ph} \times R_{ch})^{\max}$, white -- the minima of NEP. (c) Band diagrams along the BLG channel corresponding to high voltage responsivity. The respective doping voltages are indicated by color dots in panel (b). Red and black dots correspond to largest responsivity (d) Voltage responsivity obtained by scanning of the second gate voltage $V_{tg2}$ while keeping the first gate voltage fixed, at different values of the back gate voltage marked by color. Red and blue arrows indicate the photovoltage oscillations arising from the excitation of graphene plasmons. (e) Extracted maximum voltage responsivity (gray) and current responsivity (red) as a function of induced band gap. Filled and hollow symbols correspond to slightly different focusing conditions at different days of the measurement (f) Extracted noise equivalent power of the detector vs the induced band gap}
    \label{fig3}
\end{figure}

Two examples of the photovoltage maps recorded at back gate voltage $V_{bg}= V_{bg}^{\rm cnp} = -0.6 V$ and $V_{bg}= -1.42 V$ upon varying the top gate voltages $V_{tg1}$ and $V_{tg2}$ are shown in Fig.~\ref{fig3} (a) and (b). Straight vertical line at $V_{tg1}\approx 3$ V and a straight horizontal line at $V_{tg2} \approx 3$ V correspond to the charge neutrality conditions at the respective gates. The change in polarity under these gates results in the change of photovoltage sign. The photovoltage is maximized in the second and fourth quadrants of the map, which correspond to the opposite doping of the regions under the top gates. In the first and third quadrants of the map, the photovoltage is generally lower due to similar doping of the gated regions ($n-n$ in the first quadrant and $p-p$ in the third quadrant). The photovoltage is close to zero on the diagonal of the map $V_{tg1} = V_{tg2}$, as the latter corresponds to the fully uniform device doping.

The six-fold sign change of the photovoltage on the passage of the map is the hallmark of the thermoelectric mechanism of photovoltage formation~\cite{Gabor_thermoelectric}. We have verified that the overall sign of the photovoltage also agrees with the thermoelectric picture. Indeed, the photovoltage measured at the drain contact is negative for induced $p-n$ junction and positive for induced $n-p$ junction. It corresponds to the thermal diffusion of electrons from the hot mid-channel to the cold drain in the $p-n$ configuration with accumulation of negative charge at the drain terminal, and vice versa for $n-p$ junction and hot hole diffusion. The role of junctions in the vicinity of the source and drain terminals is expected to be minor due to their proximity to the metal leads and consequently low carrier temperatures.

At given back gate voltage, the absolute maxima of photovoltage are located in either second or in the fourth quadrant. The junction configurations favoring the photovoltage maximization correspond to weak doping under one gate and strong opposite doping at another one ($n^- - p^+$, $n^+ - p^-$, $p^+ - n^-$ and $p^- - n^+$ configurations). The points of the map corresponding maximum signal $V^{\max}_{\rm ph}$ are shown with red dots in Fig.~\ref{fig3}, (a, b). We also marked the points of maximum photocurrent $I^{\max}_{\rm ph}$ (cyan dots), and minimum noise equivalent power (white crosses). The positions of the $V_{\rm ph}$ maxima cannot be explained by the maximization of Seebeck coefficient difference of the two gated regions $S_1 - S_2$. The latter is maximized for weak and opposite doping of the two regions ($n^- - p^-$ and $p^- - n^-$ configurations), i.e. in the immediate vicinity of the map center. 

The explanation of the displaced optimum comes, in part, from the maximization of electric power dissipated in the channel once its resistance is low. Indeed, for fixed voltage $V_{\rm ant}$ developed by the THz antenna, the dissipated power scales as $P_{ch} = V_{\rm ant}^2/R_{ch}$, and grows for large doping and low channel resistance $R_{ch}$. The positions of maximum photovoltage per dissipated power $(V_{\rm ph} \times R_{ch})^{\max}$ are shown with yellow dots in Fig.~\ref{fig3} a,b, and indeed appear in the necessary map regions. However, the normalization of photovoltage per dissipated power is not universal, and does not explain the position of maxima at largest $E_g$. This calls for advanced model of thermoelectric voltage in such device configurations, accounting for non-uniform heating along the channel and the presence of relatively wide area in the mid-channel with fixed doping (see energy band diagram in Fig.~\ref{fig3}, c).

We have recorded the photovoltage maps upon variation of the back gate voltage, which corresponds to the induction of variable band gaps in the BLG channel. The absolute values of photovoltage and photocurrent grow throughout the map with increasing $V_{\rm bg}$, and so do the maximum values $V^{\max}_{\rm ph}$. The maximum responsivity $r^{\max}_v =V^{\max}_{\rm ph} / P$, where $P$ is the power incident on antenna, is growing with increasing $V_{\rm bg}$ (and $E_g$) both in positive and negative directions. The extracted dependence $r^{\max}_v(E_g)$ is shown in Fig.~\ref{fig3} (e) with the black line. The growth in voltage responsivity at large gaps is almost linear, and no traces of saturation are achieved even at the largest gaps $E_g \approx 90$ meV. %The bend in the responsivity curve in the top left corner in Fig.~\ref{fig3} (e) corresponds to the maximum oscillations of the photoresponse arising from the excitation of graphene plasmons, so we expect the same oscillations in the maximum photosignal with the band gap. 
More interestingly, the current responsivity $r^{\max}_I = r^{\max}_V/R_{ch}$ also displays no saturation. The noise equivalent power, limited by the thermal Johnson noise and is estimated as ${\rm NEP} =\sqrt{4 kT R_{ch}}/r_V$, does not saturate as well and continues to drop at largest $E_g$ (see Fig.~\ref{fig3}, f).

The optimal values of $r^{\max}_V$, $r^{\max}_I$ and ${\rm NEP}_{\min}$ are located at slightly different positions of the map $V_{\rm ph}(V_{\rm tg1},V_{\rm tg2})$ due to the variations of the channel resistance $R_{\rm ch}$ with $V_{\rm tg1}$ and $V_{\rm tg2}$. Remarkably, the optimal figures of merit are achieved at not very high values of channel resistance $R^{\rm opt}_{\rm ch} \sim 7$ k$\Omega$, which is several orders of magnitude below the resistance at charge neutrality (at the same $E_g$). The Fermi level in such optimal situation is located close to the band edge in either gated region, and the transport does not fall into the hopping regime. From applied viewpoint, the maximization of photovoltage at relatively low channel resistance has important implications for matching of the photodetector and its possible load~\cite{Sakowicz2011}.

Another interesting aspect of responsivity maximization with respect to the doping is the presence of oscillatory patterns in $r_V$ with varying $V_{\rm tg1}$ and $V_{\rm tg2}$. The patterns emerge predominantly at large doping of either gated section, precisely where the photoresponse is maximized. Such photovoltage oscillations are best visualized in line cuts of the photovoltage map, shown in Fig.~\ref{fig3} d. These oscillations are important for optimization of maximum photoresponse, and result in a downward bend of $r_v^{\max}(E_g)$ in the top right corner of Fig.~\ref{fig3} (e). No such oscillations appear in resistance $R_{\rm ch}(V_{\rm tg1},V_{\rm tg2})$ and its derivatives. This precludes their emergence from oscillatory strength of radiation coupling to graphene, i.e. excitation of graphene plasmons in the gated sections. Formation of the standing plasma waves under the gates can enhance or reduce the local THz field at the junctions, thereby affecting the photovoltage~\cite{Nikitin_acoustic,Ryzhii_Schottky}. The oscillation period $\Delta V_g \sim 200$ mV is in fair agreement with the theory of gated 2d plasmons. The achieved plasmon resonance frequency $f=130$ GHz is currently the lowest among those observed in graphene terahertz detectors~\cite{Bandurin_resonant,Notario_room_temp_plasmons,Cai_plasmonic_detection}. Previous reports of sub-THz plasmons were limited to graphene capacitors~\cite{Graef_2018} or detectors based on ultra-clean GaAs heterostructures~\cite{Muravev_detector,Muravev_interferometer}. From applied viewpoint, overlay of the plasmonic fringes with maximum responsivity is accidental, and no kind of 'plasmonic responsivty enhancement'~\cite{Dyakonov_detection} should be called for in this detector. The photoresponse is maximized in the same regions of $(V_{\rm tg1},V_{\rm tg2})$-map even when plasmonic fringes are absent (i.e. at higher temperatures). An intriguing fundamental problem is emergence of plasma fringes only at large $V_{\rm bg}$. It implies that gap induction prolongs the carrier momentum relaxation time in BLG. This problem did not yet gain proper theoretical consideration in the literature~\cite{Min_Disorder_BLG}.

Absence of responsivity saturation at largest band gaps in bilayer graphene is encouraging from applied viewpoint. Still, it poses a question about possible limiting factors for responsivity at even larger $E_g$. As for graphene bilayer, $E_g > 100$ meV can be hardly reached due to the breakdown of the gate dielectrics. Still, the question about responsivity scaling at large $E_g$ is relevant to black phosphorous~\cite{Viti_BlackP_detectors} and mercury cadmium telluride quantum wells (both have $E_g \lesssim 300$ meV). We can suggest that large tunneling resistance of the gapped $p-n$ junction can be the limiting factor to the NEP, as the Johnson noise is growing with the junction resistance. Large tunneling resistance can also limit the measured photovoltage responsivity once the device resistance becomes comparable with that of voltmeters.

To conclude, we have experimentally demonstrated linear growth in voltage and current responsivities with increasing the band gap in graphene bilayer sub-terahertz cryogenic detectors. At $T=7$ K and $f=130$ GHz, the voltage responsivity grows from $\sim 2$ kV/W at zero band gap up to $40$ kW/W at $E_g \approx 90$ meV. The current responsivity and signal-noise ratio also grow with gap induction without saturation. The noise equivalent power drops from $450$ fW/Hz$^{1/2}$ in the gapless state to $30$ fW/Hz$^{1/2}$ at large $E_g$. In the gapped state, excitation of plasmons in gated GBL sections affects the electromagnetic coupling and should be taken into account for responsivity maximization.

The work was supported by the grant \# 21-79-20225 of the Russian Science Foundation. The devices were fabricated using the
equipment of the Center of Shared Research Facilities (MIPT).

\bibliography{References}

\end{document}